\documentclass[prd,aps,twocolumn,showpacs,nofootinbib]{revtex4-1}

\usepackage{amsmath,graphicx,color,epsfig}
\usepackage{pstricks}
\usepackage{float}
\usepackage{subfigure}

\newcommand{\Vcb}{|\ensuremath{V_{\mathrm{cb}}}|}

\begin{document}

\title{$|V_{\rm cb}|$ from the semileptonic decay $B\to D \ell \bar{\nu}_{\ell}$ and the properties of the $D$-meson distribution amplitude}

\author{Hai-Bing Fu}
\author{Xing-Gang Wu}
\email{email:wuxg@cqu.edu.cn}
\author{Hua-Yong Han}
\author{Yang Ma}
\address{Department of Physics, Chongqing University, Chongqing 401331, P.R. China}

\author{Tao Zhong}
\address{Institute of High Energy Physics, Chinese Academy of Sciences, Beijing 100049, P.R. China}

\date{\today}

\begin{abstract}

The improved QCD light-cone sum rule (LCSR) provides an effective way to deal with the heavy-to-light transition form factors (TFFs). Firstly, we adopt the improved LCSR approach to deal with the $B\to D$ TFF $f^{+}(q^2)$ up to twist-4 accuracy. Due to the elimination of the most uncertain twist-3 contribution and the large suppression of the twist-4 contribution, the obtained LCSR shall provide us a good platform for testing the $D$-meson leading-twist DA. For the purpose, we suggest a new model for the $D$-meson leading-twist DA ($\phi_{3D}$), whose longitudinal behavior is dominantly determined by a parameter $B$. Moreover, we find its second Gegenbauer moment $a^D_2\sim B$. Varying $B$ within certain region, one can conveniently mimic the $D$-meson DA behavior suggested in the literature. Inversely, by comparing the estimations with the experimental data on the $D$-meson involved processes, one can get a possible range for the parameter $B$ and a determined behavior for the $D$-meson DA. Secondly, we discuss the $B\to D$ TFF at the maximum recoil region and present a detailed comparison of it with the pQCD estimation and the experimental measurements. Thirdly, by applying the LCSR on $f^{+}(q^2)$, we study the CKM matrix element $\Vcb$ together with its uncertainties by adopting two types of processes, i.e. the $B^0/\bar{B}^0$-type and the $B^{\pm}$-type. It is noted that a smaller $B \precsim 0.20$ shows a better agreement with the experimental value on $\Vcb$. For example, for the case of $B=0.00$, we obtain $|V_{cb}|(B^0/\bar{B}^0-{\rm type})=(41.28 {^{+5.68}_{-4.82}} {^{+1.13}_{-1.16}}) \times 10^{-3}$ and $|V_{cb}|(B^{\pm}-{\rm type})=(40.44 {^{+5.56}_{-4.72}}~ {^{+0.98}_{-1.00}}) \times 10^{-3}$, whose first (second) uncertainty comes from the squared average of the mentioned theoretical (experimental) uncertainties.

\end{abstract}

\pacs{12.15.Hh, 13.25.Hw, 14.40.Nd, 11.55.Hx}

\maketitle

\section {Introduction}

The Cabibbo-Kobayashi-Maskawa (CKM) quark mixing matrix~\cite{ckm} is one of the cornerstone of the electroweak sector of Standard Model (SM). An intensive study on the CKM matrix elements is helpful for testing SM and for exploring new physics beyond SM. Among those matrix elements, the precise determination of $\Vcb$ provides a stringent test of the CKM mechanism of the flavor structure and the charge-parity violation.

The CKM matrix element $\Vcb$ can be determined from inclusive and exclusive semi-leptonic $B$ meson decays. Experimental studies on $\Vcb$ has been done in the literature, cf.Refs.\cite{Babar_1,Babar_2,Babar_3,Belle_1,Belle_2,Belle_3,Cleo_1,Cleo_2,Cleo_3}. The recent world average of the particle data group (PDG) shows $\Vcb = 41.9(7)\times 10^{-3}$ from inclusive processes and $\Vcb = 39.6(9)\times10^{-3}$ from exclusive processes~\cite{PDG}. Theoretically, it has been studied by using non-perturbative approaches such as the quenched and unquenched lattice QCD~\cite{Lattice_1,Lattice_2,Lattice_3,Lattice_4}. It has also been perturbatively studied by using the operator production expansion approach~\cite{ope1,ope2}. The Heavy Flavor Averaging Group (HFAG) has performed a global analysis of the inclusive observables in $B\to X_c\ell\nu$ decays~\cite{HFAG_1,HFAG_2,HFAG_3,HFAG_4}, which results in $41.68(44)(9)(58)\times 10^{-3}$ and $42.31(36)\times 10^{-3}$ by using the kinetic scheme and the $1S$ scheme, respectively.

Among the $B$-meson decay channels, the semi-leptonic decays such as $B\to D \ell \bar{\nu}_\ell$ have aroused people's great interests, which have been frequently used to determine the value of $\Vcb$ and/or the $D$-meson distribution amplitude (DA). In dealing with those semi-leptonic channels, it is necessary to have a reliable estimation on the $B\to D$ transition form factors (TFFs) within its allowable kinematic region $0 \leq {q^2} \leq (M_B-M_D)^2$. Because the charm quark mass is much smaller than the bottom quark mass, the $B \to D$ TFFs can be calculated by using the light-cone sum rule (LCSR) for proper kinematical range in which the operator product expansion can be done near the light cone. More over, similar to $B\to\pi$ TFFs, e.g. Ref.\cite{bpi}, it is reasonable to assume that the $B\to D$ TFFs can also be consistently analyzed via the pQCD, the lattice QCD and the LCSR approaches, which are applicable within different $q^2$ regions. The pQCD is applicable for smaller $q^2\sim0$ and the lattice QCD is applicable for larger $q^2\sim (M_B-M_D)^2$. While, the LCSR, in which the non-perturbative dynamics are effectively parameterized in the so-called light-cone distribution amplitudes (LCDAs)~\cite{lcsr0,lcsr1,lcsr2,lcsr3}, is restricted to small and moderate $q^2$ regions. Thus, a more accurate LCSR shall present a better connection between the pQCD and the lattice QCD estimations, and then to achieve a better understanding of those TFFs.

In the present paper, we shall calculate the $B\to D$ TFFs by using the LCSR approach, and determine $\Vcb$ by comparing with the experimental values on the decay width of the $B\to D$ semileptonic decays. For the purpose, we shall analyze two types of semi-leptonic channels: $B^0 \to D^- \ell^+ \nu_\ell$ $(\bar B^0\to D^+ \ell^- \bar\nu_\ell)$ and $B^+ \to \bar D^0 \ell^+ \nu_\ell$ $(B^- \to D^0 \ell^- \bar\nu_\ell)$. In the LCSR, a two-point correlation function is introduced and expanded near the light cone $x^2=0$, whose matrix elements are parameterized as LCDAs of increasing twists~\cite{lcsr0,lcsr1,lcsr2,lcsr3}. We shall adopt the LCSR with its improved version~\cite{huangbpi1,huangbpi2} to deal with the process. By using the improved LCSR, a chiral current correlator is taken as the starting point such that the relevant (most uncertain) twist-3 LCDAs make no contributions and the reliability of LCSR estimation can be enhanced to a large degree. Further more, we shall show that the twist-4 DAs also have quite small contributions to the LCSR. This inversely makes the $B\to D$ semi-leptonic decays be good places for testing different models of the $D$ meson leading-twist LCDA.

For the purpose, we shall introduce a model for the $D$-meson leading-twist DA based on the well-known Brodsky-Huang-Lepage (BHL) prescription~\cite{BHL_1}, whose longitudinal behavior is dominantly determined by an input parameter $B$. Varying $B$ within certain region, one can conveniently mimic the $D$-meson DA behavior suggested in the literature. A comparison of three typical $D$-meson DA models shall also be presented. By comparing the estimations with the experimental data on the $D$-meson involved processes, one may get a possible range for the parameter $B$ and a determined behavior for the $D$-meson DA.

The remaining parts of the paper are organized as follows. In Sec.II, we present the calculation technologies for dealing with the dominant components of $B\to D$ semi-leptonic processes: I) We present the detail for deriving the dominant TFF $f_+(q^2)$ within the LCSR up to twist-4 accuracy; II) We present several models for $D$ meson leading twist wave function (and hence its LCDA); III) The $B$ and $D$ decay constants are two important physical quantities for determining the $B\to D$ TFFs. We present the $B$ meson and $D$ meson decay constants up to the next-to-leading order (NLO) level. Numerical results are given in Sec.III, where the properties of the $D$ meson LCDA, the $B\to D$ TFFs and the $\Vcb$ are presented. The final section is reserved for a summary.

\section{Calculation technology}

The $B\to D$ TFFs relevant for the semi-leptonic decay $B \to D \ell \bar{\nu}_\ell$ can be parameterized as
\begin{widetext}
\begin{eqnarray}
\langle D(p_D)|\bar{c}\gamma_{\mu}b|B(p_B)\rangle &=& \left[ p_{B\mu} + p_{D\mu} - \frac{m_B^2-m_D^2}{q^2}q_{\mu} \right] f_+(q^2) + \frac{m_B^2-m_D^2}{q^2} q_{\mu} f_0(q^2) \nonumber \\
&=& 2f_+(q^2)p_{D\mu} + [f_+(q^2 )+f_-(q^2)]q_{\mu}, \label{eq:matrix1}
\end{eqnarray}
\end{widetext}
where the momentum transfer $q=(p_B - p_D)$ and the relation between $f_0(q^2)$ and $f_{\pm}(q^2)$, i.e. $f_0(q^2)=f_+(q^2)+{q^2}/{(m_B^2-m_D^2)}f_{-}(q^2)$, has been adopted. After integrating over the phase space, the differential decay width of $B{\to}D\ell \bar{\nu}_\ell$ over $q^2$ can be written as
\begin{widetext}
\begin{eqnarray}
\frac{d}{dq^2}\Gamma(B{\to}D\ell\bar{\nu}_\ell) = \frac{{G_F^2 \Vcb^2 }}{192\pi^3 m_B^3} \left(1 - \frac{m_\ell^2}{q^2} \right)^2 \left[ \left( {1 + \frac{{m_\ell^2 }}{2q^2}} \right)\lambda ^{\frac{3}{2}} (q^2) |f_{+}(q^2 )|^2 + \frac{{3m_\ell^2 }}{{2q^2 }}(m_B^2  - m_D^2 )^2 \lambda ^{\frac{1}{2}} (q^2 ) |f_{0}(q^2)|^2 \right] , \label{Vcb1}
\end{eqnarray}
\end{widetext}
where $G_F=1.166\times10^{-5} \;{\rm GeV^{-2}}$ is the Fermi constant and $\lambda(q^{2}) = (m_B^2  + m_D^2  - q^2)^2 - 4 m_B^2 m_D^2$ is the phase-space factor. For the case of $\ell = e$ or $\mu$, $m_{\ell}\to 0$, the term involving $f_0(q^2)$ shall play a negligible role. This is the so-called chiral suppression. More specifically, by taking the limit $m_{\ell}\to 0$, we have
\begin{equation}
\frac{d\Gamma}{dq^2}(B{\to}D\ell\bar{\nu}_\ell) =\frac{G_F^2\Vcb^2} {192 \pi^3m_B^3} \lambda^{\frac{3}{2}}(q^2)|f_{+}(q^2)|^2 .  \label{eq:width1}
\end{equation}
The TFF $f_{+}(q^2)$ is an important component for the semi-leptonic decay and has been calculated by the lattice QCD approach~\cite{Lattice_1}, the pQCD approach~\cite{H_N_Li_1} and the QCD LCSR approach~\cite{ZuoFen_1}. If we have known the TFF $f_{+}(q^2)$ well, one can extract $\Vcb$ by comparing with the data, i.e. via the following equation
\begin{equation}
\frac{{\cal B}(B{\to}D\ell\bar{\nu}_\ell)}{\tau(B)} = \int_{0}^{(m_B  - m_D )^2 } {dq^2 \frac{{d\Gamma (B{\to}D\ell\bar{\nu}_\ell)}}{{dq^2 }}} . \label{eq:width2}
\end{equation}
Here $\tau(B)$ stands for the $B$ meson lifetime and ${\cal B}(B{\to}D\ell\bar{\nu}_\ell)$ stands for the branching ratio of $B{\to}D\ell\bar{\nu}_\ell$, both of which are experimentally measurable parameters.

\subsection{LCSR for the TFF $f_{+}(q^2)$}

For the $B$ meson decays to light meson, its basic quantity for a LCSR calculation is the correlation function of the weak current and a current evaluated between the vacuum and a light meson. To figure out the dominant twist-2 contribution and make it a better platform for determining the properties of the twist-2 LCDA, we adopt the following chiral correlation function (i.e. the correlator) to do our calculation,
\begin{widetext}
\begin{eqnarray}
\Pi_\mu(p_D,q)&=& i\int d^4x e^{ipx} \langle D(p_D)|{\rm {T}}\{\bar{c}(x) \gamma_{\mu} (1+\gamma_5) b(x),\bar{b}(0)i(1+\gamma_5)d(0)\}|0\rangle \nonumber\\
&=& \Pi(q^2,(p_{D}+q)^2) p_{D\mu} + \bar{\Pi}(q^2,(p_{D} +q)^2)q_\mu .  \label{eq:cc}
\end{eqnarray}
\end{widetext}
In stead of using the current $\bar{b} i\gamma_5 d$ for the pseudoscalar $B$ meson, we adopt a chiral current $\bar{b} i(1+\gamma_5) d$ as firstly suggested in Ref.\cite{huangbpi1} to do our calculation. The advantage of such a choice lies in that the contributions from the twist-3 LCDAs are eliminated exactly due to chiral correlation suppression. This treatment is at the price of introducing an extra contribution from a scalar $B$ meson with $J^{P}=0^+$ corresponding to operator $\bar{b} d$. To suppress the error caused by such treatment, one can set the continuum threshold parameter $s_0$ to be the one close to the lowest scalar $B$ meson, which is smaller than the pseudoscalar $B$ meson mass. This is the reason why for the improved LCSR approach the value of $s_0$ is usually taken to be those lower than the conventional LCSR. An uncertainty analysis on the choice of $s_0$ shall be presented in our numerical estimations.

On the one hand, for large (negative) virtualities of those currents, the correlator in the coordinate-space is dominated by distances close to the light-cone ($x^2\sim0$) and can be treated within the framework of light-cone expansion. On the other hand, the same correlator can be written as a dispersion relation, in the virtuality of the current coupling to $B$ meson. Equating the light-cone expansion with the dispersion relation, and separating the lowest lying $B$ meson contribution from those of higher states via quark-hadron duality, one obtains the required LCSR for the TFFs describing $B\to$ light meson decays. In this way, the LCSR allows the calculation of the properties of nonexcited hadron-states with a reasonable theoretical uncertainty. Following such standard procedures, we can obtain the LCSR for $f_{+}(q^2)$. To shorten the paper, we only list the main results and also the new results from the $D$-meson twist-4 terms, the interesting readers may turn to Ref.\cite{ZuoFen_1} for detailed calculation technology.

Up to twist-4 accuracy, the QCD LCSR for $f_{+}(q^2)$ can be written as
\begin{widetext}
\begin{eqnarray}
{f^ + }({q^2}) &=& \frac{{m_b^2{f_D}}}{{m_B^2{f_B}}}{e^{m_B^2/{M^2}}}\left\{ {\int_\Delta ^1 {du} \exp \left[ { - \frac{{m_b^2 - \bar u({q^2} - um_D^2)}}{{u{M^2}}}} \right]\left[ {\frac{\phi_D(u)}{u}} \right.} \right.\left. { - \frac{{8m_b^2[{g_1}(u) + {G_2}(u)]}}{{{u^3}{M^4}}} + \frac{{2{g_2}(u)}}{{u{M^2}}}} \right]\nonumber\\
&&\left. { + \int_0^1 {dv} \int D {\alpha _i}\frac{{\theta (\xi  - \Delta )}}{{{\xi ^2}{M^2}}}\exp \left[ { - \frac{{m_b^2 - \bar \xi ({q^2} - \xi m_D^2)}}{{\xi {M^2}}}} \right]\left[ {2{\varphi _ \bot }({\alpha _i}) + 2{{\tilde \varphi }_ \bot }({\alpha _i}) - {\varphi _\parallel }({\alpha _i}) - {{\tilde \varphi }_\parallel }({\alpha _i})} \right]} \right\} , \label{basicfq2}
\end{eqnarray}
\end{widetext}
in which $\bar{u} = 1 - u$, $\xi = \alpha_1 + v \alpha_3$, $\bar \xi = 1 - \xi$, $G_2(u)=\int_{0}^{u}g_{2}(v)dv $ and the integration upper limit is
\begin{eqnarray}
\Delta &=& \frac{1}{2m_D^2}\Big[\sqrt{(s_0-q^2-m_D^2)^2+4m_D^2(m_b^2-q^2)} \nonumber\\
&& \quad\quad\quad -(s_0-q^2-m_D^2)\Big] .
\end{eqnarray}
In addition to the leading-twist DA $\phi_D(u)$, we need to introduce two two-particle and four three-particle twist-4 DAs, which, similar to the kaonic case with $SU_f(3)$-breaking effect, can be expressed as~\cite{pballsum2}
\begin{widetext}
\begin{eqnarray}
{g_1}(u) &=& \frac{{\bar uu}}{6}[ - 5\bar uu(9{h_{00}} + 3{h_{01}} - 6{h_{10}} + 4\bar u{h_{01}}u + 10\bar u{h_{10}}u) + {a_{10}}(6 + \bar uu(9 + 80\bar uu))] \nonumber\\
&&+ {a_{10}}{{\bar u}^3}(10 - 15\bar u + 6{{\bar u}^2})\ln \bar u + {a_{10}}{u^3}(10 - 15u + 6{u^2})\ln u, \nonumber\\
{g_2}(u) &=& \frac{{5\bar uu(u - \bar u)}}{2}[4{h_{00}} + 8{a_{10}}\bar uu - {h_{10}}(1 + 5\bar uu) + 2{h_{01}}(1 - \bar uu)].\label{twist4_2particle}
\end{eqnarray}
\begin{eqnarray}
\varphi_{\perp}(\alpha_i) &=& 30 \alpha_3^2(\alpha_2-\alpha_1)[
h_{00}+h_{01}\alpha_3+\frac{1}{2}\,h_{10}(5\alpha_3-3)] , \nonumber\\
\widetilde{\varphi}_{\perp}(\alpha_i) &=& -30 \alpha_3^2[
h_{00}(1-\alpha_3)+h_{01}\Big[\alpha_3(1-\alpha_3)-6\alpha_1\alpha_2\Big] + h_{10}\Big[\alpha_3(1-\alpha_3)-\frac{3}{2}(\alpha_1^2
+\alpha_2^2)\Big]],\nonumber\\
{\varphi}_{\parallel}(\alpha_i) &=& 120 \alpha_1\alpha_2\alpha_3
[ a_{10} (\alpha_1-\alpha_2)],\nonumber\\
\tilde{\varphi}_{\parallel}(\alpha_i) &=& 120 \alpha_1\alpha_2
\alpha_3 [ v_{00} + v_{10} (3\alpha_3-1)],\label{twist4_3particle}
\end{eqnarray}
\end{widetext}
where
\begin{eqnarray}
h_{00} & = & v_{00} =-\frac{\delta^2}{3}, \;  a_{10} = \delta^2\epsilon-\frac{9}{20} a^D_2 m_D^2,\nonumber\\
v_{10} & = & \delta^2\epsilon, \; h_{01} = \frac{2}{3}\delta^2\epsilon-\frac{3}{20} a^D_2 m_D^2 \nonumber
\end{eqnarray}
and
\begin{displaymath}
h_{10} = \frac{4}{3}\delta^2\epsilon+\frac{3}{20} a_2 m_D^2 .
\end{displaymath}
Here, as a rough estimation of $D$-meson twist-4 contributions, we adopt  $\delta^2(1{\rm GeV})=0.20 {\rm GeV}^2$ and $\epsilon(1{\rm GeV})=0.53$~\cite{pballsum2}. The uncertainties for such approximation are suppressed by the fact that the twist-4 part itself contributes less than $4\%$ of the total TFF, which will be shown latter discussions.

Taking the limit of infinite quark masses, our present TFF $f^+(q^2)$ coincides with the Isgur-Wise function for the TFFs between heavy mesons~\cite{Isgur1,Isgur2}. This shows that at least at the leading order level, the LCSR for $f^+(q^2)$ are equivalent to the estimations by taking the heavy quark symmetry. At the NLO level, the heavy quark mass effect may cause changes among those two approaches, which is out of the range of the present paper. In order to conveniently compare with the experimental analysis done in the literature, we also present LCSR for the $B\to D$ TFF within the heavy quark symmetry.

The non-perturbative matrix element defined in Eq.(\ref{eq:matrix1}) can be treated by taking the heavy quark limit, which shall result in the following form,
\begin{eqnarray}
&& \langle D(p_D)|\bar{c}\gamma_{\mu}b|B(p_B)\rangle  = \nonumber\\
&&  \sqrt{m_B~m_D}[h_+(w)(v_B+v_D)_\mu+h_-(w)(v_B-v_D)_\mu] , \label{eq:ffv}
\end{eqnarray}
where the four velocities $v_B={p_B}/{m_B}$ and $v_D={p_D}/{m_D}$. The relationship between $f_{+}(q^2)$ and $h_{+ (-)}(w)$ is
\begin{equation}
f_{+}(q^2)=\frac{m_B+m_D}{2\sqrt{m_B~m_D}}{\cal G}(w), \label{eq:f12}
\end{equation}
where
\begin{displaymath}
{\cal G}(w)=h_+(w)-\frac{m_B-m_D}{m_B+m_D}h_-(w)
\end{displaymath}
and $w$ stands for the product of $B$ meson and $D$ meson four velocities, which is defined as
\begin{equation}
w=v_B\cdot v_D=\frac{m^2_B + m^2_D-q^2}{2 m_B m_D} .  \label{wrelation}
\end{equation}
When $q^2\to 0$, we get its maximum value, $w_{max}=(m^2_B + m^2_D)/(2 m_B m_D)$. When $q^2\to (m_B -m_D)^2$, we get its minimum value, $w_{min}=1$.

Then, the LCSR for TFF ${\cal G}(w)$ takes the following form
\begin{widetext}
\begin{eqnarray}
{\cal G}(w) &=& \frac{{2m_b^2m_D^{1/2}}}{{({m_B} + {m_D})m_B^{3/2}}}\frac{{{f_D}}}{{{f_B}}}{e^{m_B^2/{M^2}}}\bigg\{ \int_\Delta ^1 {du} \exp \bigg[ { - \frac{{m_b^2 - \bar u(m_B^2 + \bar um_D^2 - 2{m_B}{m_D}w)}}{{u{M^2}}}} \bigg]\left[ {\frac{{{\phi _D}(u)}}{u}}\right. \nonumber\\
&&\left. { - \frac{{8m_b^2[{g_1}(u) + {G_2}(u)]}}{{{u^3}{M^4}}} + \frac{{2{g_2}(u)}}{{u{M^2}}}} \right] + \int_0^1 {dv} \int D {\alpha _i}\frac{{\theta (\xi  - \Delta )}}{{{\xi ^2}{M^2}}}\exp \left[ { - \frac{{m_b^2 - \bar \xi (m_B^2 + \bar \xi m_D^2 - 2{m_B}{m_D}w)}}{{\xi {M^2}}}} \right]\nonumber\\
&& { \times \left[ {2{\varphi _ \bot }({\alpha _i}) + 2{{\tilde \varphi }_ \bot }({\alpha _i}) - {\varphi _\parallel }({\alpha _i}) - {{\tilde \varphi }_\parallel }({\alpha _i})} \right]} \bigg\} . \label{LCSRgw}
\end{eqnarray}
\end{widetext}

\subsection{Models for the leading-twist $D$ meson DA}

The leading-twist $D$ meson DA has the asymptotic form, $\phi^{\rm as}_D(x,\mu^2)|_{\mu\to\infty} = 6x\bar x$. In practical applications, we need to know what is the shape of $D$ meson DA at low and moderate energy scales. The DA at any scale $\mu$ can be expanded in Gegenbauer series as
\begin{eqnarray}
\phi_D(x, \mu^2) = 6x\bar x \sum^\infty_{n=0} a^D_n(\mu^2) C^{3/2}_n(x-\bar x),
\label{DA_Gegenbauer}
\end{eqnarray}
where $C^{3/2}_n(x-\bar x)$ are Gegenbauer polynomials and $a^D_n(\mu^2)$ are Gegenbauer moments. If the DA shape at a scale $\mu_0$ is known, we can inversely get its Gegenbauer moments by using the orthogonality relation for Gegenbauer polynomial, i.e.,
\begin{eqnarray}
a^D_n(\mu^2_0) = \frac{\int^1_0 dx \; \phi_D(x,\mu^2_0) C^{3/2}_n(x-\bar x)}
{\int^1_0 dx \; 6x\bar x[C^{3/2}_n(x-\bar x)]^2}. \label{Gegenbauer_moment}
\end{eqnarray}
Then, by including the QCD evolution effect, the $D$ meson DA at any scale can be written as~\cite{TFFas}
\begin{eqnarray}
{\phi_D}(x,\mu^2) = 6x\bar x\sum\limits_{n = 0}^\infty  {{a^D_n}
(\mu_0^2)} {\left( {\ln \frac{{{\mu^2}}}{{\Lambda _{QCD}^2}}}
\right)^{ - {\gamma_n}}}C_n^{3/2}(x - \bar x)\nonumber\\
\end{eqnarray}

As a pQCD estimation for $B\to D$ decays, by introducing a free parameter $C_d$, Ref.\cite{H_N_Li_1} has suggested a naive model for $D$ meson DA, i.e.
\begin{eqnarray}
\phi_{1D}(x)=6x(1-x)[1+C_d(x-\bar x)] .
\end{eqnarray}
By setting $C_d=0.7$, they predicte $\Vcb=0.035\sim0.036$; or inversely, if taking $\Vcb=0.04$, they predict $C_d=0.4\sim0.5$. A larger value $C_d=0.8$ has also been suggested in Ref.\cite{lucd}. In our calculation, we shall adopt $\phi_{1D}(x)$ as the first DA model to do our discussion.

On the other hand, the $D$ meson DA can be related to its light-cone wave function (LCWF) $\psi_D(x,\mathbf{k}_\bot)$ via the relation,
\begin{eqnarray}\label{DA_WF}
\phi_D (x,\mu_0^2) = \frac{{2\sqrt 6 }}{{f_D }} \int_{|{\bf{k}}_\bot|^2 \le \mu_0 ^2 } {\frac{{d{\bf{k}}_ \bot  }}{{16\pi ^3 }} \psi_D (x,{\bf{k}}_\bot )},
\end{eqnarray}
where $f_{D}$ is decay constant. Thus one could first construct a reasonable model for the $D$ meson WF and then get its DA. A proper way of constructing the $D$ meson WF/DA with a better end-point behavior at small $x$ and $k_\perp$ region is very important for dealing with high energy processes, especially for pQCD calculations.

As suggested, one useful way for modeling the hadronic valence WF is to use approximate bound-state solution of a hadron in terms of the quark model as the starting point. The BHL prescription~\cite{BHL_1} of the hadronic WF is rightly obtained via this way by connecting the equal-time WF in the rest frame and the WF in the infinite momentum frame. It shows that the longitudinal and transverse distributions for the WF $\psi_D(x,\textbf{k}_\bot)$ are entangled with each other, which can be constructed as
\begin{eqnarray}
&&\psi_D(x,\textbf{k}_\bot) \nonumber\\
&&= A_D \varphi_D(x) \exp \left[ - b_D ^2 \left(\frac{{\bf k}_\bot^2 + m_c^2 }{x} + \frac{{\bf k}_\bot^2 + m_d^2}{\bar{x}} \right)\right] , \label{WF1}
\end{eqnarray}
where $A_D$ is the overall normalization constant. For the $x$-dependent part, similar to the pionic case~\cite{XGWU_1}, we can assume $\varphi_D(x) = [1 + B \times C^{3/2}_2(x - \bar x)]$, in which $B$ is the phenomenological parameter to be fixed by studying the $D$ meson involved processes. In the following, we shall show that the value of $B$ is close to the second Gegenbauer moment, $B\sim a_{2}^D$, which basically determines the broadness of the longitudinal distribution. More over, because $m_c \gg m_d$, we shall have a large non-zero first Gegenbauer moment $a_{1}^D$ as suggested in Refs.\cite{H_N_Li_1,lucd}.

After integrating out the transverse momentum, we get the second model for the $D$ meson DA, i.e.,
\begin{widetext}
\begin{equation}
\phi _{2D} (x, \mu_0^2) = \frac{{\sqrt 6 A_{2D} x\bar{x}}} {{8\pi ^2 f_D b_{2D}^2 }}[1+B\times C_2^{3/2}(x-\bar x)] \exp \left[ { - b_{2D}^2
\frac{{x m_d^2  + \bar{x}m_c^2 }}{{x\bar{x}}}} \right]\left[ 1 - \exp \left( - \frac{{b_{2D}^{2} \mu_0^2 }}{{x\bar{x}}} \right) \right] , \label{phi2d}
\end{equation}
\end{widetext}
where the constitute light quark mass $m_d\sim300$ MeV from the constitute quark model~\cite{cqm}, $A_{2D}$ and $b_{2D}$ are undetermined parameters.

As a further step, we include the spin-space WF $\chi_D(x,\textbf{k}_\bot)= (\bar{x}m_c+xm_d)/{\sqrt{\mathbf{k}^2_\perp+(\bar{x}m_c+x m_d)^2}}$~\cite{Melosh_1}, into the WF, i.e.
\begin{eqnarray}
\psi'_D(x,\textbf{k}_\bot) = \chi_D(x,\textbf{k}_\bot)
\psi_D(x,\textbf{k}_\bot). \label{WF2}
\end{eqnarray}
Such spin-space part comes from the Wigner-Melosh rotation~\cite{Melosh_2}, whose idea is reasonable: when one transforms from equal-time (instant-form) WF to LCWF, besides the momentum space WF transformation, one should also consider the Melosh transformation relating equal-time spin WFs and light-cone spin WFs. After integrating it over the transverse momentum dependence, we get the third model for the $D$ meson DA,
\begin{widetext}
\begin{eqnarray}
\phi_{3D}(x,\mu_0^2 )&& =\frac{A_{3D} \sqrt{6x\bar{x}} Y}
{{8\pi ^{3/2} f_D b_{3D} }}[1+B\times C_2^{3/2}(x-\bar x)] \exp \bigg[  - b_{3D} ^2\frac{{x m_d^2  + \bar{x}m_c^2  - {\rm{Y}}^2 }}{{x\bar{x}}}\bigg] \left[ {{\rm{Erf}}\Big( {\frac{{b_{3D}\sqrt{
{\mu_0 ^2  + {\rm{Y}}^2 } }}}{{\sqrt {x\bar{x}} }}} \Big)
- {\rm{Erf}}\Big( {\frac{{b_{3D} {\rm{Y}}}}{{\sqrt {x\bar{x}} }}}
\Big)} \right],\nonumber\\ \label{phi3d}
\end{eqnarray}
\end{widetext}
where $A_{3D}$, $B$ and $b_{3D}$ are undetermined parameters. The error function ${\rm{Erf}}(x)$ is defined as $ {\rm{Erf}} (x)=2\int^x_0{\exp({-t^2})dt} /\sqrt{\pi} $, ${\rm{Y}} = x m_d+\bar{x}m_c$ and $\bar{x} = 1 - x $.

As for the second and third WFs, we have two constraints to determine the WF parameters:
\begin{itemize}
\item The first one is the WF normalization condition
\begin{eqnarray}
\int^1_0 dx \int \frac{d^2 \textbf{k}_\bot}{16\pi^3} \psi_D(x,\textbf{k}_\bot) = \frac{f_D}{2\sqrt{6}} \;. \label{Pd_1}
\end{eqnarray}

\item The second one is the probability of finding the leading valence-quark state in $D$ meson ($P_D$), which is $\simeq 0.8$~\cite{PD_1,PD_2,sunyanjun}. Here the probability $P_D$ is defined as
    \begin{eqnarray}
     P_D=\int^1_0{dx \int_{|\mathbf{k}_{\perp}|^2\le \mu_0^2} {\frac{d^2 \mathbf{k}_\perp}{16\pi^3}|\psi_D(x, \mathbf{k}_{\perp} )|^2}}.
    \end{eqnarray}
    More specifically, for the above mentioned WF models (\ref{WF1},\ref{WF2}), we obtain
    \begin{widetext}
    \begin{eqnarray}
     {P_{2D}}&=& \frac{{A_{2D}^2}}{{32{\pi ^2}b_{2D}^2}}\int_0^1 {dx} \varphi^2(x)x\bar x\exp \left[ { - 2b_{2D}^2\frac{{m_d^2x + m_c^2\bar x}}{{x\bar x}}} \right] \left[ {1 - \exp \left( { - 2b_{2D}^2\frac{{\mu _0^2}}{{x\bar x}}} \right)} \right] , \\
     P_{3D}  &=& \frac{{A_{3D} ^2 }}{{16\pi ^2 }}\int_0^1 {dx}\varphi^2(x)
     {\rm{Y}}^2 \exp \left[ { - 2 b_{3D}^2 \frac{{m_d^2 x + m_c^2\bar x}}{{x\bar x}}}  \right] \int_0^{\mu_0 ^2 } {\frac{{dk_ \bot ^2 }}{{k_ \bot ^2  + {\rm{Y}}^2 }}} \exp \left( { - 2b_{3D} ^2 \frac{{k_ \bot ^2 }}{{x\bar x}}} \right) .
    \end{eqnarray}
   \end{widetext}

\end{itemize}

The remaining free parameter $B$ can be fixed by comparing with the data, and then the WF/DA behavior can be determined finally. In combination with the above two constraints, it is noted that by using a proper value of $B$, most of the DA shapes suggested in the literature can be simulated.

\subsection{Decay constants for the $B$ and $D$ mesons}

The $B$ and $D$ decay constants are two important physical quantities for determining the $B\to D$ TFFs and the $D$ meson DA.

A comparative study on the $B$ meson decay constant under several different correlation functions has been done in Ref.\cite{1002.0483}. To be consistent with our present LCSR analysis on the $B\to D$ TFF, we adopt the chiral correlation function to do the calculation, i.e.
\begin{displaymath}
\Pi (q^2 ) = i\int {d^4 x} e^{iq \cdot x} \langle 0|\bar q(x)(1 + \gamma _5 )b(x),\bar b(0)(1 - \gamma _5 )q(0)|0\rangle .
\end{displaymath}
Following the standard procedure, we can obtain the sum rule for $f_{B}$ up to NLO,
\begin{eqnarray}
f_B^2&&\frac{{m_B^4}}{{m_b^2}}{e^{ - m_B^2/{M^2}}} = \nonumber\\
&&\frac{3}{{4{\pi ^2}}}\int_{m_b^2}^{{s_0}} {ds\,s\,{e^{ - s/{M^2}}}} {(1 - x)^2}\left[ {1 + \frac{{{\alpha _s}({\mu _{\rm IR}}){C_F}}}{\pi }\rho (x)} \right]\nonumber\\
&&+ {e^{ - m_b^2/{M^2}}}\bigg[ {\frac{1}{6}\langle \frac{{{\alpha _s}}}{\pi }GG\rangle  - \frac{{32\pi }}{{27}}\frac{{{\alpha _s}({\mu _{\rm IR}}){{\langle \bar qq\rangle }^2}}}{{{M^2}}}} \nonumber\\
&&\times {\bigg( {1 - \frac{{m_b^2}}{{4{M^2}}} - \frac{{m_b^4}}{{12{M^4}}}} \bigg)} \bigg] , \label{fbsr}
\end{eqnarray}
where $m_b$ stands for the $b$-quark, $\mu_{\rm IR}$ is the renormalization scale, $x=m_b^2/s$ and $C_F=4/3$. The parameters $M$ and $s_0$ stand for the Borel parameter and the effective continuum threshold respectively. The function $\rho(x)$ determines the spectral density of the NLO correction to the perturbative part,
\begin{eqnarray}
\rho (x) =&& \frac{9}{4} + 2{{\rm Li}_2}(x) + \ln x\ln (1 - x) - \ln (1 - x)\nonumber\\
&&+ \left( {x - \frac{3}{2}} \right)\ln \frac{{1 - x}}{x} - \frac{x}{{1 - x}}\ln x ,  \label{rho1}
\end{eqnarray}
where ${\rm{Li}}_2(x)$ means the dilogarithm function. Practically, $\rho(x)$ is firstly derived under the $\overline{MS}$ scheme, and then transformed into Eq.(\ref{rho1}) with the help of the well-known one loop formula for the relation between the $b$ quark $\overline{MS}$-mass and the pole mass, i.e.
\[
\overline{m}_b(\mu_{\rm{IR}})=m_b\left[1+ \frac{ \alpha_s(\mu_{\rm{IR}}) C_F}{4\pi} \left(-4+3\ln\frac{m_b^2} {\mu_{\rm IR}^2}\right)\right].
\]

By changing all $B$ meson parameters to the corresponding $D$ meson parameters, we can get similar LCSR as Eq.(\ref{fbsr}) for $f_D$.

\section{Numerical results and discussions}

\subsection{Input parameters}

As for the heavy quark masses, we take $m_b=4.85\pm0.05$ GeV and $m_c=1.50\pm0.05$ GeV. For $B$ and $D$ mesons' masses, we take $m_B=5.279$ GeV and $m_D=1.869$ GeV~\cite{PDG}. We take the light condensates $\langle\bar{q}q\rangle$ and $\langle\frac{\alpha_s}{\pi}G_{\mu\nu}^aG^{a\mu\nu}\rangle$ as~\cite{duplan,cond}
\begin{eqnarray}
\langle\bar{q}q\rangle(1\;{\rm GeV}) &=& -(0.246^{+0.018}_{-0.019}\;{\rm GeV})^3\nonumber\\
\langle\frac{\alpha_s}{\pi}GG\rangle &=& 0.012^{+0.006}_{-0.012}\;{\rm GeV}^4\nonumber\\
\langle\bar{q}g\sigma\cdot Gq\rangle(1\;{\rm GeV}) &=& (0.8\pm0.2)\; {\rm GeV}^2\langle\bar{q}q\rangle(1\;{\rm GeV}), \nonumber
\end{eqnarray}
where $q$ denotes light $u$ or $d$ quark.

\subsection{The $B$ and $D$ decay constants}

\begin{table}[htb]
\begin{center}
\begin{tabular}{c| c c c }
\hline\hline
~~$m_b/{\rm GeV}$~~ & ~~$s_0/{\rm GeV}^2$~~ & ~~$M^2/{\rm GeV}^2$~~ & ~~$f_B/{\rm GeV}$~~ \\
\hline
$4.80$ & $[32.8, 35.9]$ & $[1.93, 2.36]$ & $0.160(5)$ \\
\hline
$4.85$ & $[32.5, 34.9]$ & $[1.81, 2.17]$ & $0.141(4)$ \\
\hline
$4.90$ & $[32.3, 33.9]$ & $[1.84, 2.00]$ & $0.121(2)$ \\
\hline \hline
\end{tabular}
\caption{The $B$ meson decay constant $f_B$ up to NLO for $m_b=4.85\pm0.05$ GeV. The number in the parenthesis shows the uncertainty in the last digit. }
\label{tabp1}
\end{center}
\end{table}

\begin{table}[htb]
\begin{center}
\begin{tabular}{c| c c c }
\hline\hline
~~$m_c/{\rm GeV}$~~ & ~~$s_0/{\rm GeV}^2$~~ & ~~$M^2/{\rm GeV}^2$~~ & ~~$f_D/{\rm GeV}$~~ \\
\hline
$1.45$ & $[5.07, 5.95]$ & $[0.67, 0.81]$ & $0.180(5)$ \\
\hline
$1.50$ & $[5.31, 5.72]$ & $[0.59, 0.73]$ & $0.163(4)$ \\
\hline
$1.55$ & $[4.92,5.01]$  & $[0.67, 0.68]~$ & $0.142(6)$ \\
\hline \hline
\end{tabular}
\caption{The $D$ meson decay constant $f_D$ up to NLO for $m_c=1.50\pm0.05$ GeV. The number in the parenthesis shows the uncertainty in the last digit. }
\label{tab_fd}
\end{center}
\end{table}

\begin{figure}[htbp]
\begin{center}
\includegraphics[width=0.5\textwidth]{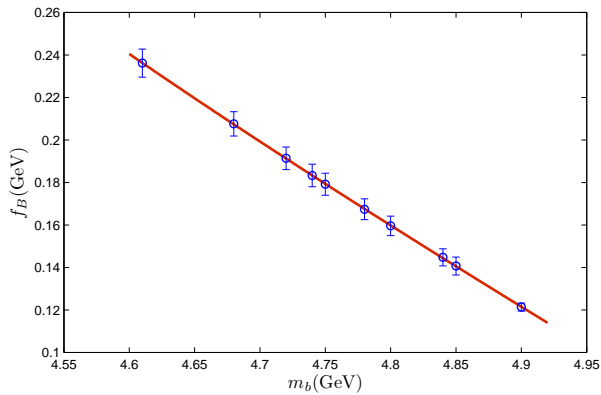}
\end{center}
\caption{The value $f_B$ versus $m_b\in[4.60,~4.90]$ GeV, in which the errors for some specific points are caused by the choices of $s_0$ and the Borel window within their allowable regions. } \label{fb}
\end{figure}

The $B$ and $D$ decay constants are usually studied via their leptonic decay channels, earlier discussions of which can be found in Ref.\cite{khlopov}. At present, we determine the $B$ and $D$ decay constants from the LCSR (\ref{fbsr}). The Borel window, i.e. the allowable range of the Borel parameter $M^2$, and the effective continuous threshold $s_0$ can be determined from three restriction conditions: I) The continuum contribution is not higher than 30\%; II) The dimension-six condensate does not exceed 15\%; III) The estimated $B$ meson mass compared with the experimental results does not exceed 1\%. A LCSR for $m_B$ can be easily derived by doing the derivative of the logarithm of Eq.(\ref{fbsr}) with respect to $1/M^2$, which can be conveniently adopted for determining the $B$ meson mass. The results are presented in Tables \ref{tabp1} and \ref{tab_fd}.

Tables \ref{tabp1} and \ref{tab_fd} indicate that the value of $f_B$ or $f_D$ decreases almost linearly with the increment of $b$ or $c$ quark mass. This can be seen by Fig.(\ref{fb}), which represents the behavior of $f_B$ versus $m_b$. Here the errors are caused by varying $s_0$ within the region listed in Table \ref{tabp1} and by varying $M^2$ within the allowable Borel window. In the literature, based on the non-relativistic constituent quark model or via an application of the Dyson-Schwinger equation, it has been known that $f_{B}|_{m_b\to\infty}\propto 1/\sqrt{m_B}$~\cite{fb0,fb1,fb2,fb3,fb4}. On the other hand, under the QCD sum rule approach, such asymptotic behavior shall be altered by a certain degree when we have taken the non-perturbative terms proportional to the quark and gluon condensates into consideration~\cite{fb5}. A similar linear $m_b$ dependence has also been observed in a recent QCD sum rule analysis~\cite{fb6}.

\subsection{The $D$ meson distribution amplitude}

\begin{table}[tb]
\begin{center}
\begin{tabular}{c| c c c c c}
\hline
\hline
~~$m_c$~~ & ~~$\mu_0$~~ & $~~A_{3D}~~$ & $~~b_{3D}~~$ & $~~A_{2D}~~$ & $~~b_{2D}~~$ \\
\hline
                             & 1 & 416.6 & 0.791 & 514.8 & 0.841\\
\raisebox {2.0ex}[0pt]{1.45} & 2 & 479.9 & 0.812 & 595.8 & 0.862\\
\hline
                             & 1 & 739.9 & 0.854 & 937.2 & 0.902\\
\raisebox {2.0ex}[0pt]{1.50} & 2 & 814.1 & 0.868 & 1033  & 0.915\\
\hline
                             & 1 & 1674.  & 0.940 & 2184.  & 0.985\\
\raisebox {2.0ex}[0pt]{1.55} & 2 & 1763.  & 0.947 & 2301.  & 0.991\\
\hline
\hline
\end{tabular}
\caption{The WF parameters $A_{2D}$, $b_{2D}$, $A_{3D}$ and $b_{3D}$ with $m_c=1.50\pm0.05$ GeV. $B=0.00$. The value of $f_D$ is taken as the central value for each $m_c$ and we adopt two initial scales for DA, i.e. $\mu_0=1$ and $2$ GeV respectively. }
\label{tabp3}
\end{center}
\end{table}

\begin{figure}[tb]
\begin{center}
\includegraphics[width=0.5\textwidth]{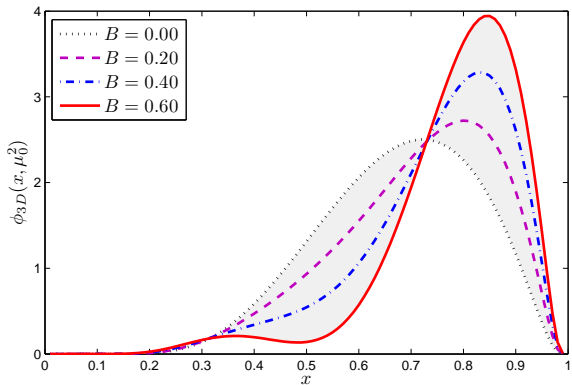}
\end{center}
\caption{The $D$ meson DA $\phi_{3D}(x,\mu^2_{0})$ at $\mu_0=1$ GeV with different $B$, in which we have set $B=0.00$,$\cdots$,$0.60$, respectively. } \label{DA_2}
\end{figure}

\begin{table}[htb]
\begin{center}
\begin{tabular}{c | c | c | c}
\hline
\hline
~~Model~~ & ~~$B$~~ & ~~$a_1^D(\mu^2_0 =1\;{\rm GeV}^2)$~~ & ~~$a_2^D(\mu^2_0 =1\;{\rm GeV}^2)$~~ \\
\hline
    & 0.00 & 0.625 & 0.056\\
    & 0.10 & 0.618 & 0.135\\
II  & 0.20 & 0.614 & 0.211\\
    & 0.30 & 0.612 & 0.289\\
    & 0.40 & 0.611 & 0.370\\
\hline
    & 0.00 & 0.586 & 0.024\\
    & 0.10 & 0.581 & 0.103 \\
III & 0.20 & 0.576 & 0.180\\
    & 0.30 & 0.579 & 0.258\\
    & 0.40 & 0.576 & 0.341\\
\hline
\end{tabular}
\caption{The first and second Gegenbauer moments of the $D$ meson leading-twist DAs $\phi_{2D}(x,\mu_0^2)$ and $\phi_{3D}(x,\mu_0^2)$ for typical $B$ within the region of $[0.00,0.40]$. $m_d=0.30$ GeV, $m_c=1.50$ GeV, $P_D=0.8$ and $\mu_0=1$ GeV. }
\label{Gegenbauer_moment}
\end{center}
\end{table}

As a combination of the above mentioned two constraints, i.e. the normalization condition (\ref{Pd_1}) and the probability $P_D=0.8$, we determine the $D$ meson DA parameters. We put the results for the DA parameters $A_{2D}$, $b_{2D}$, $A_{3D}$ and $b_{3D}$ in Table \ref{tabp3}, where we have set $B=0.00$ as an explicit example and all other parameters are set to be their central values. During the calculation, the parameter $B$ could be treated as a free parameter for determining the DA models $\phi_{2D}$ and $\phi_{3D}$. We put the $D$ meson DA $\phi_{3D}$ with different choices of $B$ in Fig.(\ref{DA_2}), in which we have set the value of $B$ up to a larger value of $0.60$. It is found that by varying $B$ within a certain region, e.g. $B\in[0, 0.6]$, the $D$ meson DA shall vary from asymptotic-like to double-humps-like, then, one reproduces most of the $D$ meson DAs suggested in the literature. This agrees with our experience on pion DA~\cite{XGWU_1}. Then, inversely, by comparing the estimations with the experimental data on $D$ meson involved processes, one can obtain the possible range for the parameter $B$ and then a determined behavior of the $D$ meson DA.

The first and second Gegenbauer moments $a_1^D(1{\rm GeV}^2)$ and $a_2^D(1{\rm GeV}^2)$ with varying $B\in[0.00,0.40]$ for $\phi_{2D}$ and $\phi_{3D}$ are presented in Table \ref{Gegenbauer_moment}. By setting $B\in[0.00,0.40]$, we get the steady first Gegenbauer moment, i.e. $a^{D}_{1}\sim[0.61,0.63]$ for $\phi_{2D}$ and $a^{D}_{1}\sim[0.57,0.59]$ for $\phi_{3D}$. These vales are consistent with those of Ref.\cite{H_N_Li_1}, which, at present, is a natural deduction of our present LCDA model. More over, we observe that the value of the second Gegenbauer moment $a^{D}_{2}\sim B$, which shows that the parameter $B$ does basically determine the broadness of the longitudinal distribution of $D$ meson DA.

\begin{figure}[tb]
\begin{center}
\includegraphics[width=0.5\textwidth]{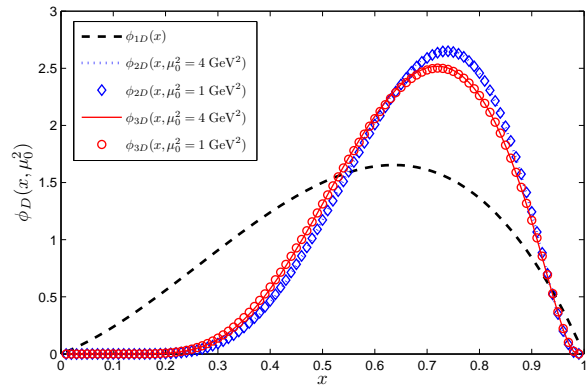}
\end{center}
\caption{Three $D$ meson DAs $\phi_{1D,2D,3D}$ under two different scales, in which we have set $B=0.00$ for $\phi_{2D}$ and $\phi_{3D}$. The curves for $\phi_{2D}$ or $\phi_{3D}$ at the two scales are almost coincide with each other. As for $\phi_{1D}$, we set $C_d=0.70$~\cite{H_N_Li_1}. } \label{DA}
\end{figure}

As a comparison, we present the $D$ meson DAs $\phi_{1D,2D,3D}(x,\mu^2_0)$ in Fig.(\ref{DA}). It shows that the $D$ meson DA shape changes slightly by varying the scale $\mu_0$ from $1$ GeV to $2$ GeV. And as a comparison of $\phi_{2D}$ and $\phi_{3D}$, by including the spin-space WF effect, the DA end-point behavior can be further improved.

\subsection{The $B\to D$ transition form factor}

Using the QCD LCSR for the $B\to D$ TFF $f^+(q^2)$, we discuss its properties in detail. The TFF $f^+(q^2)$ or ${\cal G}(w)$ depends weakly on the allowable Borel window $M^2\in[15,19]\;{\rm GeV}^{2}$, and we shall fix $M^2$ to be $17\;{\rm GeV}^2$ to do our calculation.

\begin{figure}[htb]
\begin{center}
\includegraphics[width=0.5\textwidth]{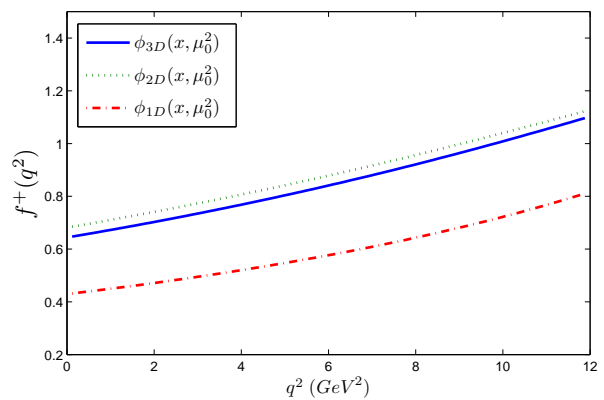}
\end{center}
\caption{The TFF $f^+(q^2)$ for three $D$ meson DAs. The dash-dot, the dotted and the solid lines are for $\phi_{1D}$, $\phi_{2D}$ and $\phi_{3D}$, respectively. For the case of $\phi_{2D}$ and $\phi_{3D}$, we have set $B=0.00$ and $\mu_0=1$ GeV. } \label{fq2phi}
\end{figure}

We present the TFF $f^+(q^2)$ up to twist-4 accuracy for the $D$ meson DAs $\phi_{1D,2D,3D}$ in Fig.(\ref{fq2phi}). The shapes/trends of the three curves are similar to each other. The simplest model $\phi_{1D}$, which agrees with that of Ref.\cite{ZuoFen_1} by using the same inputs, provides much lower $f^+(q^2)$ in the whole $q^2$ region than those of $\phi_{2D}$ and $\phi_{3D}$. Thus, the previously adopted naive DA model $\phi_{1D}$ can only provide the conceptional estimation on $f^+(q^2)$. The TFF $f^+(q^2)$ for both $\phi_{2D}$ and $\phi_{3D}$ are close to each other. This is reasonable, since the TFFs are dominated by large $x$ region that is close to $1$ and $\phi_{2D}$ and $\phi_{3D}$ have similar behaviors in this region. The inclusion of spin-space WF shall lead to a more accurate estimation, so, we take $\phi_{3D}(x,\mu_0)$ as the $D$ meson DA to do our following discussions.

\begin{figure}[tb]
\begin{center}
\includegraphics[width=0.5\textwidth]{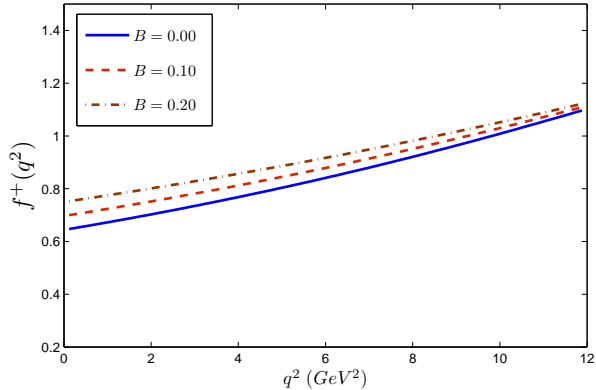}
\end{center}
\caption{The TFF $f^+(q^2)$ for $D$ meson DA $\phi_{3D}$ with different choice of $B$. The solid, the dashed and the dash-dot lines are for $B=0.00$, $0.10$ and $0.20$, respectively. } \label{fq2phi3B}
\end{figure}

The TFFs for $\phi_{3D}$ with $B=0.00$, $0.10$ and $0.20$ are presented in Fig.(\ref{fq2phi3B}). It shows that $f^+(q^2)$ increases with the increment of $B$. This agree with the trends shown in Fig.(\ref{DA_2}) that a bigger $B$ leads to a weaker suppression in the end-point region ($x\to 0$ or $x\to 1$), and shall result in a larger estimation on $f^{+}(q^2)$.

\begin{figure}[tb]
\begin{center}
\includegraphics[width=0.5\textwidth]{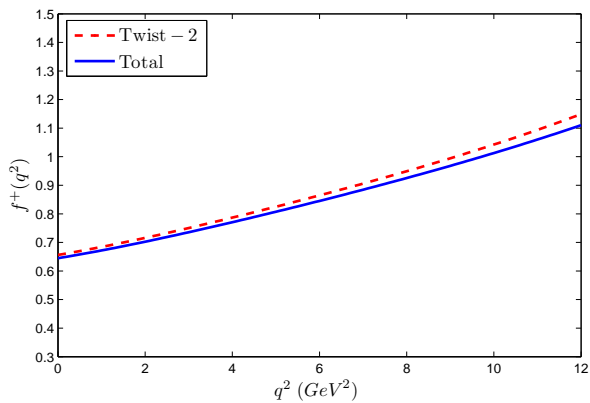}
\end{center}
\caption{The TFF $f^+(q^2)$ for $\phi_{3D}$ with $B=0.00$ up to twist-4 accuracy. It shows the twist-2 part provides dominant contribution, while the twist-4 part gives quite small negative contribution. }\label{fq2}
\end{figure}

To compare the relative importance of different twist structures, we present the TFF $f^+(q^2)$ for the twist-2 part only and the total TFF up to twist-4 accuracy in Fig.(\ref{fq2}), where the $D$ meson DA is taken as $\phi_{3D}$ with $B=0.00$. The cases for other $B$ values are similar. As required, Fig.(\ref{fq2}) shows that the twist-2 part provides dominant contribution, while the twist-4 part gives quite small (negative) contribution. The twist-4 contribution slightly increases with the increment of $q^2$, and for $q^2=12\;{\rm GeV}^2$, the twist-4 part provides $\sim 4\%$ absolute contribution to the TFF $f^+(12)$. The twist-4 part should be taken into consideration in cases when a physical observable sizably depends on the TFF at large $q^2$ region. Fig.(\ref{fq2}) also indicates that our present treatment of $D$-meson twist-4 DAs is viable, since the twist-4 DAs for $D$ meson and kaon are similar (both are treated as heavy-and-light meson) and their differences to the total TFF, and hence to the following determined $\Vcb$, can be highly suppressed by the total quite small twist-4 contributions to the integrated TFF in whole $q^2$ region.

\begin{table}[tb]
\begin{center}
\begin{tabular}{|c |c |c |c| }
\hline
~~Refs.~~ & ~~\cite{Lattice_1}~~ & ~~\cite{Lattice_2}~~ & ~~\cite{Lattice_3}~~ \\
\hline
${\cal G}(1)$ & 1.026(17) & 1.074(24) & 1.058(20) \\
\hline
\end{tabular}
\caption{The value of TFF ${\cal G}(\omega)$ at the minimum recoil point, ${\cal G}(1)$, under the quenched lattice QCD approach~\cite{Lattice_1,Lattice_2,Lattice_3}, where the number in parenthesis shows its uncertainty in the last digit.} \label{G1-tab}
\end{center}
\end{table}

In the literature, one always uses ${\cal G}(w)$ for pQCD and experimental analysis, especially for determining the CKM matrix element $\Vcb$, cf.Refs.\cite{Belle_1,Cleo_1,Cleo_2,Cleo_3}. An important input for the experimental fit is ${\cal G}(w=1)$, which is the value of TFF at the minimum recoil point (corresponding to $q^2=(m_{B}-m_{D})^2$). Theoretically, we have $h_+(1)\to 1$ and $h_-(1)\to 0$ in the framework of the heavy quark effective theory, which results in the limiting behavior ${\cal G}(1)\to 1$. The quenched lattice QCD estimation~\cite{Lattice_1,Lattice_2,Lattice_3}, cf. Table \ref{G1-tab}, shows ${\cal G}(1)\to 1$ could be a good approximation.

\begin{figure}[tb]
\begin{center}
\includegraphics[width=0.5\textwidth]{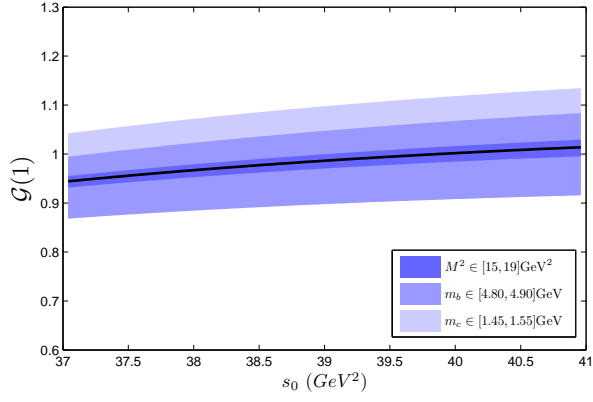}
\end{center}
\caption{The TFF ${\cal G}(1)$ by varying $s_0$ within the wide region of $[37,41]\;{\rm GeV}^2$, where the uncertainties for $m_c\in[1.45,1.55]$ GeV, $m_b\in[4.80,4.90]$ GeV and $M^2\in[15,19]\;{\rm GeV}^2$ are presented by shaded bands, respectively. The central solid line is for $m_c=1.50$ GeV, $m_b=4.85$ GeV and $M^2=17\;{\rm GeV}^2$. }\label{G1}
\end{figure}

\begin{figure}[tb]
\begin{center}
\includegraphics[width=0.5\textwidth]{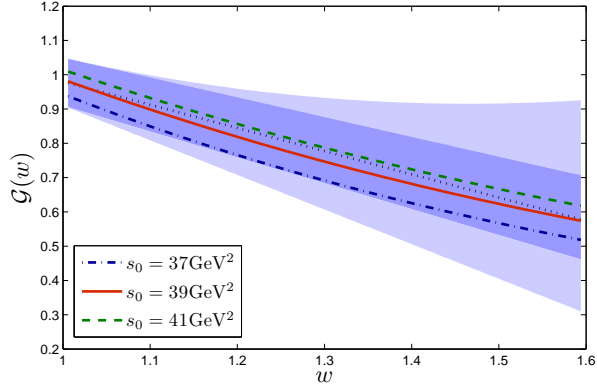}
\end{center}
\caption{The QCD LCSR for TFF ${\cal G}(w)$ versus $w$, in which the allowable range for $w$ is $[1.00,1.59]$. The dash-dot, the solid and the dashed lines are for $s_0=37\;{\rm GeV}^2$, $39\;{\rm GeV}^2$ and $41\;{\rm GeV}^2$, respectively. As a comparison, we also present the parametrization (\ref{bellefit}) of Belle Collaboration~\cite{Belle_1}: the dotted line is the central value for ${\hat{\rho}}^2 _D=0.69$ and ${\hat c_D=0.00}$, the thicker shaded band shows the uncertainty of linear fit and the lighter shaded band is for quadratic fit.  }   \label{Gw_1}
\end{figure}

Using the LCSR (\ref{LCSRgw}), we put our prediction of ${\cal G}(1)$ versus the threshold parameter $s_0$ in Fig.(\ref{G1}), where the uncertainties for $m_c\in[1.45,1.55]$ GeV, $m_b\in[4.80,4.90]$ GeV and $M^2\in[15,19]\;{\rm GeV}^2$ are presented. Our central value is $[0.94,1.01]$ for $s_0\in[37,41]\;{\rm GeV}^2$. The value of ${\cal G}(1)$ is steady over the Borel window, which changes by less than $2\%$ for $M^2\in[15,19]\;{\rm GeV}^2$. Varying $w$ within its allowable range of $[1.00,1.59]$, the TFF ${\cal G}(w)$ for several continuum threshold $s_0$ is drawn in Fig.(\ref{Gw_1}). By varying $s_0$ within the wide region from $37\;{\rm GeV}^2$ to $41\;{\rm GeV}^2$, ${\cal G}(1)$ changes from $\pm7\%$ to $\pm8\%$ for $m_b\in[4.80,4.90]$ GeV and from $\left(^{+13\%}_{-6\%}\right)$ to $\left(^{+14\%}_{-7\%}\right)$ for $m_c\in[1.45,1.55]$ GeV, respectively. As a combined (squared) error for the $b$ and $c$ quark mass uncertainties, it is found that ${\cal G}(1)$ changes by $\left(^{+15\%}_{-9\%}\right)$ at $s_0=37\;{\rm GeV}^2$ and $\left(^{+16\%}_{-11\%}\right)$ at $s_0=41\;{\rm GeV}^2$.

Experimentally, ${\cal G}(w)$ is usually parameterized as the following form~\cite{Cleo_1,Cleo_2,Cleo_3,Belle_1}:
\begin{eqnarray}
{\cal G}_D(w) &=& {\cal G}_D(1)\left[ 1 - {\hat{\rho}} _D^2(w - 1) + {\hat c_D}{(w - 1)^2} \right. \nonumber\\
&& \quad\quad\quad \left. + {\cal O}((w - 1)^3) \right], \label{bellefit}
\end{eqnarray}
in which the undetermined parameters are taken as~\cite{Belle_1}
\begin{eqnarray}
{\hat{\rho}}^2_D &=& 0.69 \pm 0.14,\quad {\hat c_D} = 0.00
\end{eqnarray}
for the linear fit; and
\begin{eqnarray}
{\hat{\rho}}^2_D &=& 0.69^{+0.42}_{-0.15},\quad {\hat c_D} = 0.00 ^{+0.59}_{-0.00}
\end{eqnarray}
for the quadratic fit. As a comparison of our theoretical estimations, we have also put the results for the parametrization (\ref{bellefit}) in Fig.(\ref{Gw_1}): the dotted line is the central value for ${\hat{\rho}}^2 _D=0.69$ and ${\hat c_D=0.00}$, the lighter shaded band is the uncertainty of quadratic fit and the thicker shaded band is for linear fit. Fig.(\ref{Gw_1}) shows our present prediction of ${\cal G}(w)$ is in a good agreement with the data, which also consistent with the pQCD estimation at the large recoil region~\cite{H_N_Li_1}.

\subsection{The matrix element $\Vcb$ and its uncertainties}

There are four $B\to D$ semi-leptonic processes that are frequently used to determine the CKM matrix element $\Vcb$, i.e. $B^0 \to D^- \ell^+ \nu_\ell$ and $\bar B^0\to D^+ \ell^- \bar\nu_\ell$, $B^+ \to \bar D^0 \ell^+ \nu_\ell$ and $B^- \to D^0 \ell^- \bar\nu_\ell$. The branching ratios and lifetimes of those processes can be grouped into two types, one is called as the ``$B^0/\bar{B}^0$-type" with~\cite{PDG}
\begin{eqnarray}
{\cal{B}}(B^0 \to D^- \ell^+ \nu_\ell) &=& {\cal{B}}(\bar B^0\to D^+ \ell^- \bar\nu_\ell ) \nonumber\\
&=& (2.18\pm0.12)\% ,\nonumber\\
\tau (B^0\;{\rm or}\;\bar{B}^0)&=&1.519\pm0.007 \;{\rm ps} ,\nonumber
\end{eqnarray}
and the other is called as the ``$B^{\pm}$-type" with~\cite{PDG}
\begin{eqnarray}
{\cal{B}}(B^+ \to \bar D^0 \ell^+ \nu_\ell) &=& {\cal{B}}(B^-\to D^0 \ell^- \bar\nu_\ell ) \nonumber\\
&=& (2.26\pm0.11)\% , \nonumber\\
\tau(B^\pm)&=&1.641\pm0.008 \;{\rm ps} . \nonumber
\end{eqnarray}
In the following, we shall adopt those two types of processes to determine $\Vcb$.

\begin{table}[t]
\begin{center}
\begin{tabular}{|c | c| c | c| c|}
\hline\hline
$(B^0\to D^-\ell^ + \nu_{\ell})$ & ~~$|V_{cb}^{\rm Max}|$~~ & ~~$\Delta^+$~~  &  ~~$|V_{cb}^{\rm Min}|$~~ &  ~~$\Delta^-$~~ \\
\hline
$m_b=(4.85\pm0.05)$ GeV  & 44.74 & +3.45 & 37.41 & -3.88 \\
\hline
$m_c=(1.50\pm0.05)$ GeV  & 43.66 & +2.37 & 40.01 & -1.28 \\
\hline
$s_0=(39 \pm 2)\;{\rm GeV}^2$  & 44.96 & +3.68 & 38.87 & -2.41 \\
\hline
$M^2=(17 \pm 2)\;{\rm GeV}^2$  & 42.36 & +1.08 & 40.43 & -0.86 \\
\hline\hline
${\cal B}=(2.18\pm0.12)\%$  & 42.40 & +1.12 & 40.13 & -1.15 \\
\hline
$\tau=(1.519 \pm 0.007)$ ps & 41.38 & +0.10 & 41.19 & -0.10 \\
\hline\hline
\end{tabular}
\caption{Theoretical and experimental uncertainties for $|V_{cb}|$ under the $B^0/\bar{B}^0$-type. The central value is $|V_{cb}^{\rm CV}|=41.28$, which is obtained by setting all parameters to be their cental values. The symbols CV, Max and Min stand for the central value, the maximum value and the minimum value, respectively. The conditions for the $B^{\pm}$-type are similar. }
\label{uncertainty_1}
\end{center}
\end{table}

\begin{figure}[ht]
\begin{center}
\includegraphics[width=0.45\textwidth]{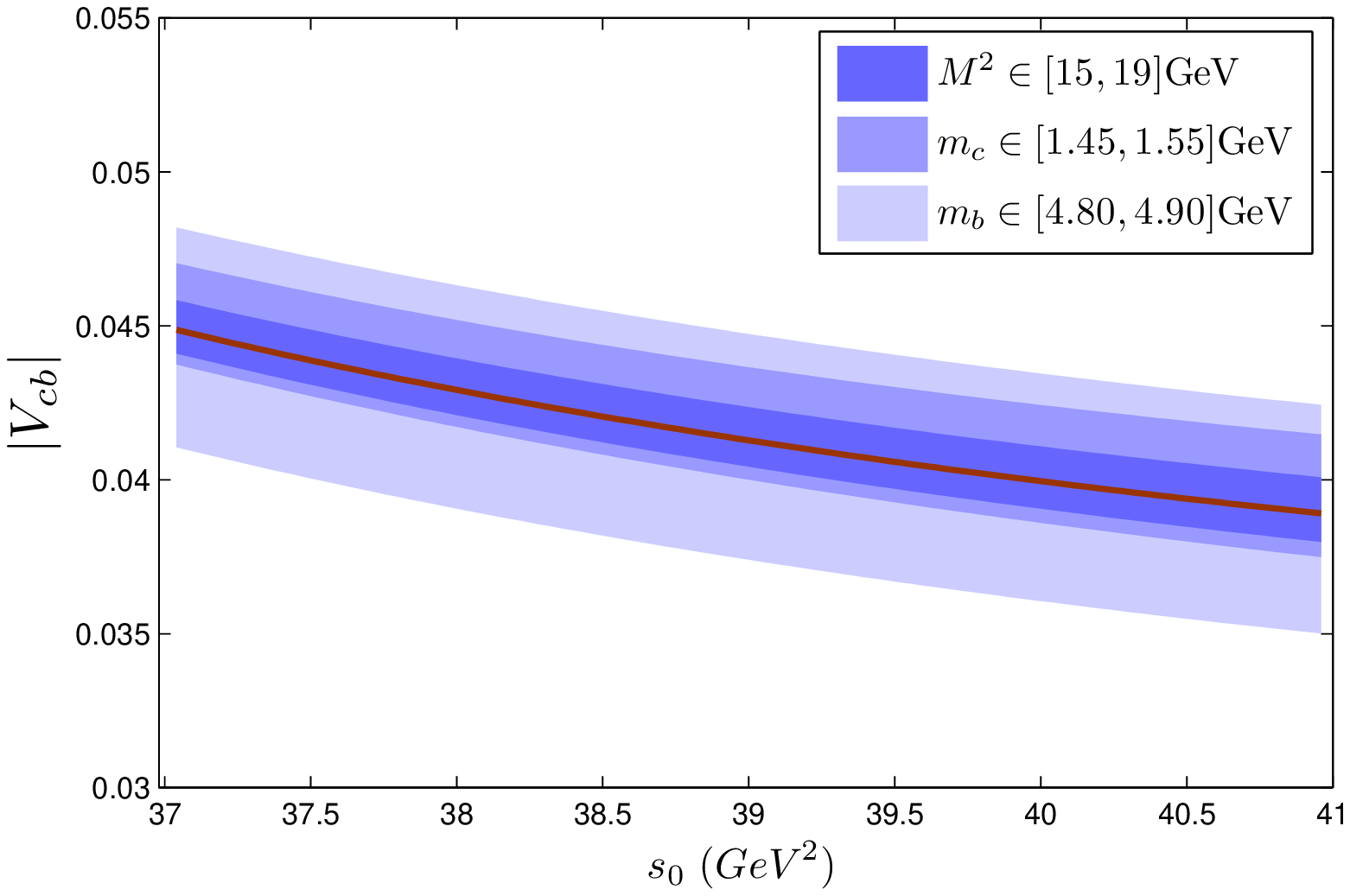}
\includegraphics[width=0.45\textwidth]{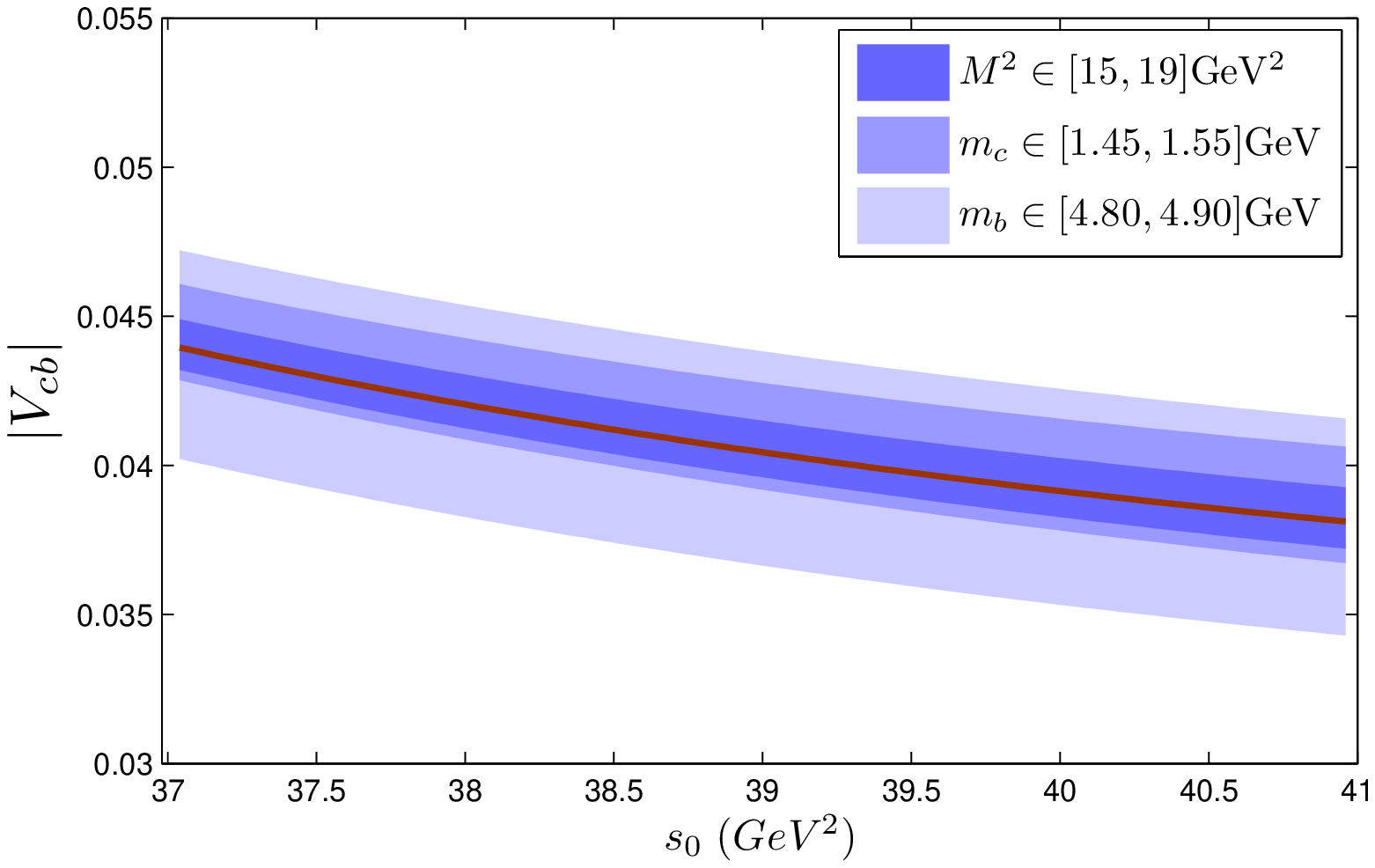}
\end{center}
\caption{The uncertainties of $\Vcb$ versus $s_0$ from the QCD LCSR analysis, where the left is for $B^0/\bar{B}^0$-type and the right is for $B^{\pm}$-type. The shaded bands are for uncertainties of different parameters, which are derived by varying these parameters within their reasonable regions as $m_b=(4.85\pm0.05)$ GeV, $m_c=(1.50\pm0.05)$ GeV and $M^2=(17 \pm 2)\;{\rm GeV}^2$. The solid line stands for the central values of $\Vcb$. } \label{GBDw}
\end{figure}

\begin{table}[tb]
\begin{center}
\begin{tabular}{|c | c | c| }
\hline\hline
~~~$B$~~~ & ~~~~~~$B^0/\bar{B}^0$-type~~~~~~ & ~~~~~~$B^{\pm}$-type~~~~~~ \\
\hline
0.00 & $41.28 {^{+5.68}_{-4.82}}~ {^{+1.13}_{-1.16}}$  & $40.44 {^{+5.56}_{-4.72}}~ {^{+0.98}_{-1.00}}$ \\
\hline
0.10 & $39.50 {^{+5.36}_{-4.68}}~ {^{+1.08}_{-1.11}}$  & $38.70 {^{+5.25}_{-4.58}}~ {^{+0.94}_{-0.96}}$ \\
\hline
0.20 & $38.00{^{+ 5.17}_{- 4.59}}~ {^ {+ 1.04}_{- 1.06}}$  & $37.22 {^{+5.06}_{-4.49}}~ {^{+0.90}_{-0.92}}$ \\
\hline\hline
\end{tabular}
\caption{The value of $\Vcb$ in unit $10^{-3}$ with varying $B$ for $D$ meson DA. Three choices of $B$, i.e. $0.00$, $0.10$ and $0.20$, are adopted. The central values for $\Vcb$ are obtained by setting all inputs to be their central values. The errors are calculated by theoretical and experimental errors for all inputs, similar to the case of Table \ref{uncertainty_1}. }  \label{Vcb_results}
\end{center}
\end{table}

Taking $\phi_{3D}$ with $B=0.00$ as an example, we show how the considered uncertainty sources affect $\Vcb$, i.e.,
\begin{equation}
|V_{cb}|(B^0/\bar{B}^0-{\rm type})=(41.28 {^{+5.68}_{-4.82}}{^{+1.13}_{-1.16}}) \times 10^{-3}  \label{Vcb_21}
\end{equation}
and
\begin{equation}
|V_{cb}|(B^{\pm}-{\rm type})=(40.44 {^{+5.56}_{-4.72}}~ {^{+0.98}_{-1.00}}) \times 10^{-3}, \label{Vcb_22}
\end{equation}
in which the first (second) uncertainty comes from the squared average of the mentioned theoretical (experimental) uncertainties shown in Table \ref{uncertainty_1}. That is, the theoretical uncertainty mainly comes from the $c$ and $b$ quark masses, the Borel window and the choice of the threshold parameter $s_0$. The experimental uncertainty comes from the lifetime and the decay ratio of the mentioned processes. A clear description of those uncertainties are presented in Fig.(\ref{GBDw}).

Next,we discuss the variation of by taking $\phi_{3D}$ with several choices of $B=0.00$, $0.10$ and $0.20$, respectively. The results are put in Table \ref{Vcb_results}. It is noted that the value of $\Vcb$ decreases with the increment of $B$. To compare with the experimental estimations on $\Vcb$, we need a smaller $B$ and hence a smaller second Gegenbauer moment. This, in some sense, consistent with the present analysis for the pion DA, which also prefers an asymptotic behavior with small second Gegenbauer moment or small $B$ value~\cite{XGWU_1}.

\begin{table}[tb]
\begin{center}
\begin{tabular}{|c|c| }
\hline\hline
~~~  ~~~ &  ~~~~~$|V_{cb}|\times 10^{-3}$~~~~~ \\
\hline
BABAR \cite{Babar_1} (ULC) & $39.8(18)(13)$\\
\hline
BABAR \cite{Babar_1} (SSM) & $41.6(18)(14)$\\
\hline
PDG (Lattice) \cite{PDG}  &  $39.4(14)(13)$\\
\hline
CLEO \cite{Cleo_1}  & $45(6)(4)(5)$\\
\hline
Belle \cite{Belle_1} & $41.9(45)(53)$\\
\hline
QLC \cite{Lattice_1}   & $38.4(9)(42)$\\
\hline
DELPHI \cite{DELPHI}	& $41.4(12)(21)$\\
\hline
HQET \cite{HQET}   & $40(6)$\\
\hline
Our result ($B^0/\bar{B}^0$-type) &  $41.28 {^{+5.68}_{-4.82}}~ {^{+1.13}_{-1.16}}$     \\
\hline
Our result ($B^\pm$-type) &  $40.44 {^{+5.56}_{-4.72}}~ {^{+0.98}_{-1.00}}$ \\
\hline\hline
\end{tabular}
\caption{A comparison of $|V_{cb}|$ with some estimations done in the literature, in which the first and second errors are for theoretical and experimental uncertainty sources, respectively. The symbol QLC means the quenched lattice calculation and the HQEF means the heavy quark effective theory. } \label{tabp2}
\end{center}
\end{table}

As a final remark, we preset a comparison of $|V_{cb}|$ for $B=0.00$ with the present estimations done in the literature. We put such a comparison in Table \ref{tabp2}. Experimentally, the value of ${\cal G}(1)|V_{cb}|$ is determined in a combined way to short the uncertainties and the value of $|V_{cb}|$ is determined by using theoretical estimations on ${\cal G}(1)$. As for BABAR collaboration~\cite{Babar_1}, the SSM means using ${\cal G}(1)$ determined by the quenched lattice calculation based on the Step Scaling Method~\cite{Lattice_1} and the ULC means using ${\cal G}$ determined by the unquenched lattice calculation~\cite{Lattice_2}. Tables \ref{Vcb_results} and \ref{tabp2} show that our present QCD LCSR estimation on $\Vcb$ for a smaller $B$ shows a good agreement with the experimental estimates.

\section{summary}

In the present paper, by adopting several $D$ meson DA models, we have presented a detailed discussion on $B\to D$ TFF $f^{+}(q^2)$ or ${\cal G}(w)$ within the QCD LCSR approach. Based on the sum rules together with the experimental data on $B\to D$ semileptonic decays, we have analyzed the CKM matrix element $\Vcb$, in which a detailed error analysis has been presented.

We have calculated the $B\to D$ TFF up to twist-4 accuracy by using the improved QCD LCSR with chiral current. By using chiral current in the correlator, the most uncertain twist-3 contributions can be eliminated due to chiral suppression. It shows that the twist-2 part provides dominant contributions to the form factor and the twist-4 parts only give less than $4\%$ contributions in whole $q^2$ region. Thus this provides another platform for testing the properties of twist-2 DA.

We have newly suggested a convenient $D$ meson DA model (\ref{phi3d}) based on the BHL prescription together with the Wigner-Melosh rotation effect. As shown by Table \ref{Gegenbauer_moment}, its second Gegenbauer moment is dominantly determined by a parameter $B$, i.e. $a^D_2 \sim B$. The DA shapes for various $B$ are put in Fig.(\ref{DA_2}). By using a proper choice of $B$, most of the DA shapes suggested in the literature can be simulated. Then, if by comparing with the data, the value of $B$ can be fixed, the DA behavior can be determined accordingly. It is noted that to compare with the experimental result on $\Vcb$, a smaller $B \precsim 0.20$ shows a better agreement. By varying $B\in[0.00,0.20]$, its first Gegenbauer moment $a^{D}_1$ is about $[0.6,0.7]$, consistent with the pQCD suggestion~\cite{H_N_Li_1}.

The TFF $f^+(q^2)$ have been calculated by using three different $D$ meson DAs. As shown by Fig.(\ref{fq2phi}), the usual simple model $\phi_{1D}$ shall lead to smallest $f^+(q^2)$ and can only be adopted for a conceptional estimation on $f^+(q^2)$. By using $\phi_{3D}$, with a larger $B$ value, a larger $f^{+}(q^2)$ is observed, which is due to less suppression from the DA around the end-point region.

A detailed uncertainty analysis on ${\cal G}(1)$ has also been done. As shown by Fig.(\ref{Gw_1}), our present prediction of ${\cal G}(w)$ shows a better agreement with the data. The central value of ${\cal G}(1)$ is $[0.94,1.01]$ for $s_0\in[37,41]\;{\rm GeV}^2$, consistent with HQET limit ${\cal G}(1)\to 1$. The value of ${\cal G}(1)$ is steady over the Borel window, which changes by less than $2\%$ for $M^2\in[15,19]\;{\rm GeV}^2$.

\begin{figure}[tb]
\begin{center}
\includegraphics[width=0.5 \textwidth]{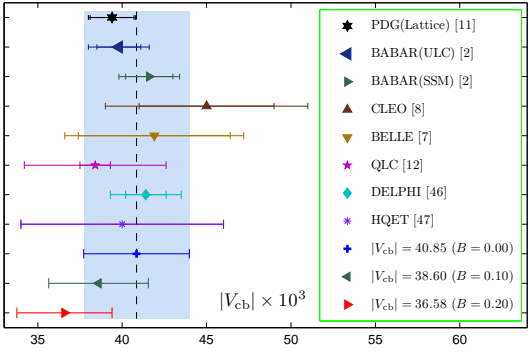}
\end{center}
\caption{A comparison of $\Vcb$ with experimental and theoretical predictions. Our estimations for $B=0.00$, $0.10$ and $0.20$ are presented. }\label{Vcb_22}
\end{figure}

The matrix element $\Vcb$ and its uncertainties have been studied by using two types of processes, e.g. the $B^0/\bar{B}^0$-type and the $B^{\pm}$-type. For the case of $B$=0, by adding the errors for all mentioned experimental and theoretical uncertainty sources, we obtain $|V_{cb}|(B^0/\bar{B}^0-{\rm type})=(41.28 {^{+6.81}_{-5.98}}) \times 10^{-3}$ and $|V_{cb}|(B^{\pm}-{\rm type})=(40.44 {^{+6.54}_{-5.72}}) \times 10^{-3}$. As a weighted average of these two types we obtain,
\begin{equation}
|V_{cb}| = (40.84\pm3.11)\times 10^{-3} , \;\;(B=0.00)
\end{equation}
where the error stands for the standard derivation of the weighted average. Similarly, we have
\begin{eqnarray}
|V_{cb}| &=& (39.08\pm3.03)\times 10^{-3} , \;\;(B=0.10) , \\
|V_{cb}| &=& (37.59\pm2.89)\times 10^{-3} , \;\;(B=0.20) .
\end{eqnarray}
A comparison of $\Vcb$ with experimental and theoretical predictions is put in Fig.(\ref{Vcb_22}), in which our estimations for $B=0.00$, $0.10$ and $0.20$ are presented. We have also shown how the considered uncertainty sources affect $\Vcb$. The results are presented in Table \ref{Vcb_results}, in which three choices of $B$ are adopted, i.e. $B=0.00$, $0.10$ and $0.20$, respectively. Through a comparison with the experimental data, our present estimation for $\Vcb$ with a small $B$ shows a good agreement with the BABAR, CLEO and Belle estimates. With more and more available data for the $D$ meson involved processes, the $D$ meson DA will be finally determined by a global fit.

\hspace{1cm}

\noindent{\bf Acknowledgments}: This work was supported in part by Natural Science Foundation of China under Grant No.11075225 and No.11275280, by the Program for New Century Excellent Talents in University under Grant No.NCET-10-0882, and by the Fundamental Research Funds for the Central Universities under Grant No.CQDXWL-2012-Z002.


\begin{thebibliography}{99}

\bibitem{ckm} N. Cabibbo, Phys. Rev. Lett. {\bf 10}, 531 (1963); M. Kobayashi and T. Maskawa, Prog. Theor. Phys. {\bf 49}, 652 (1973).

\bibitem{Babar_1} B. Aubert {\it et al.}, (BABAR Collaboration), Phys. Rev. Lett. {\bf 104}, 011802 (2010).

\bibitem{Babar_2} B. Aubert {\it et al.}, (BABAR Collaboration), Phys. Rev. D {\bf 79}, 012002, (2009).

\bibitem{Babar_3} B. Aubert {\it et al.}, (BABAR Collaboration), Phys. Rev. D {\bf 69}, 111104 (2004).

\bibitem{Belle_2} W. Dungel {\it et al.}, (Belle Collaboration), Phys. Rev. D {\bf 82}, 112007(2010).

\bibitem{Belle_3} C. Schwanda {\it et al.}, (Belle Collaboration), Phys. Rev. D {\bf78}, 032016 (2008).

\bibitem{Belle_1} K. Abe {\it et al.}, (Belle Collaboration), Phys. Lett. B {\bf 526}, 258 (2002).

\bibitem{Cleo_1} J. Bartelt {\it et al.}, (CLEO Collaboration), Phys. Rev. Lett. {\bf 82}, 3746 (1999).

\bibitem{Cleo_2} M. Athanas {\it et al.}, (CLEO Collaboration), Phys. Rev. Lett. {\bf 79}, 2208 (1997).

\bibitem{Cleo_3} D. Buskulic {\it et al.}, (ALEPH Collaboration), Phys. Lett. B {\bf 395}, 373 (1997).

\bibitem{PDG} J. Beringer {\it et al.}, (Particle Data Group), Phys. Rev. D {\bf 86}, 010001 (2012).

\bibitem{Lattice_1} G. M. de Divitiis {\it et al.}, Phys. Lett.  B {\bf 655}, 45 (2007).

\bibitem{Lattice_2} M. Okamoto {\it et al}. Nucl. Phys. Proc. Suppl. {\bf 140}, 461 (2005).

\bibitem{Lattice_3} S. Hashimoto {\it et al.}, Phys. Rev. D {\bf 61}, 014502 (1999).

\bibitem{Lattice_4} C. Bernard {\it et al.}, Phys. Rev. D {\bf 79}, 014506 (2009).

\bibitem{ope1} I. I. Bigi, M. Shifman, N. G. Uraltsev and A. Vainshtein, Phys. Rev. Lett. {\bf 71}, 496 (1993).

\bibitem{ope2} A. V. Manohar and M. B. Wise, Phys. Rev. D {\bf 49}, 1310 (1994).

\bibitem{HFAG_1} D. Benson, I. I. Bigi, T. Mannel and N. Uraltsev, Nucl. Phys. B {\bf 665}, 367 (2003).

\bibitem{HFAG_2} P. Gambino and N. Uraltsev, Eur. Phys. J. C {\bf 34}, 181 (2004).

\bibitem{HFAG_3} D. Benson, I. I. Bigi and N. Uraltsev, Nucl. Phys. B {\bf 710}, 371 (2005).

\bibitem{HFAG_4} P. Gambino and C. Schwanda, arXiv:1102.0210.

\bibitem{bpi} T. Huang and X. G. Wu, Phys. Rev. D {\bf 71}, 034018 (2005).

\bibitem{lcsr0} P. Ball, V. M. Braun and H. G. Dosch, Phys. Rev. D{\bf 44}, 3567 (1991).

\bibitem{lcsr1} I. I. Balitsky, V. M. Braun and A. V. Kolesnichenko, Nucl. Phys. B{\bf 312}, 509 (1989).

\bibitem{lcsr2} V. M. Braun and I. E. Filyanov, Z. Phys. C {\bf 44}, 157 (1989).

\bibitem{lcsr3} V. L. Chernyak and I. R. Zhitnitskii, Nucl. Phys. B {\bf 345}, 137 (1990).

\bibitem{huangbpi1} T. Huang, Z. H. Li and X. Y. Wu, Phys. Rev. D{\bf 63}, 094001 (2001).

\bibitem{huangbpi2} Z. G. Wang, M. Z. Zhou and T. Huang, Phys. Rev. D{\bf 67}, 094006 (2003).

\bibitem{BHL_1} S. J. Brodsky, T. Huang and G. P. Lepage, in Particles and Fields-2, edited by A. Z. Capri and A. N. Kamal (Plenum, New York, 1983), p.143; S. J. Brodsky, G. P. Lepage, T. Huang and P. B. MacKenzis, in Particles and Fieds 2, edited by A. Z. Capri and A. N. Kamal (Plenum, New York, 1983), p.83.

\bibitem{H_N_Li_1} T. Kurimoto, H. N. Li and A. I. Sanda, Phys. Rev. D {\bf 67}, 054028 (2003).

\bibitem{ZuoFen_1} F. Zuo, Z. H. Li and T. Huang, Phys. Lett. B {\bf 641}, 177 (2006); T. Kurimoto, H. N. Li and A. I. Sanda, Phys. Rev. D {\bf 67}, 054028 (2003); T. Huang, Z. H. Li and X. Y. Wu, Phys. Rev. D {\bf 63}, 094001 (2001).

\bibitem{pballsum2} P. Ball, JHEP {\bf 9901}, 010 (1999).

\bibitem{Isgur1} N. Isgur and M. B. Wise, Phys. Lett. B {\bf 232}, 113 (1989).

\bibitem{Isgur2} N. Isgur and M. B. Wise, Phys. Lett. B {\bf 237}, 527 (1990).

\bibitem{TFFas} G. P. Lepage and S. J. Brodsky, Phys. Rev. D {\bf 22}, 2157 (1980).

\bibitem{lucd} R. H. Li, C. D. Lu and H. Zou, Phys. Rev. D{\bf 78}, 014018 (2008).

\bibitem{XGWU_1} X. G. Wu and T. Huang, Phys. Rev. D {\bf 82}, 034024 (2010); X. G. Wu and T. Huang, Phys. Rev. D {\bf 84}, 074011 (2011); T. Huang, T. Zhong and X. G. Wu, Phys. Rev. D {\bf 88}, 034013 (2013).

\bibitem{cqm} G. Zweig, CERN Reports Th. 401 and 412, 1964; in: A. Zichichi (Ed.), Proc. Int. School of Phys. Ettore Majorana, Erice, Italy, 1964, Academic, New York, p. 192.

\bibitem{Melosh_1} T. Huang, B. Q. Ma and Q. X. Shen, Phys. Rev. D{\bf 49}, 1490 (1994); X. G. Wu, T. Huang and T. Zhong, Chin. Phys. C {\bf 37}, 063105 (2013); F. G. Cao and T. Huang, Phys. Rev. D {\bf 59}, 093004 (1999); T. Huang and X. G. Wu, Phys. Rev. D {\bf 70}, 093013 (2004); X. G. Wu and T. Huang, Int. J. Mod. Phys. A {\bf 21}, 901 (2006); X. G. Wu and T. Huang, JHEP {\bf 0804}, 043 (2008).

\bibitem{Melosh_2} E. Wigner, Ann. Math. {\bf 40}, 149 (1939); H.J. Melosh, Phys. Rev. D{\bf 9}, 1095 (1974).

\bibitem{PD_1} X. Q. Li, Z. Q. Zhang and T. Huang, Z. Phys. C {\bf 42}, 99 (1989).

\bibitem{PD_2} X. H. Guo and T. Huang, Phys. Rev. D {\bf 43}, 2931 (1991).

\bibitem{sunyanjun} Y. J. Sun, X. G. Wu, F. Zuo and T. Huang, Eur. Phys. J. C {\bf 67}, 117 (2010).

\bibitem{1002.0483} X. G. Wu, Y. Yu, G. Chen and H. Y. Han, Commun. Theor. Phys. {\bf 55}, 635 (2011).

\bibitem{duplan} G. Duplancic, A. Khodjamirian, Th. Mannel, B. Melic and N. Offen, JHEP {\bf 0804}, 014 (2008).

\bibitem{cond} B. L. Ioffe, Prog. Part. Nucl. Phys. {\bf 56}, 232 (2006).

\bibitem{khlopov} S.S.Gershtein and M.Yu.Khlopov, JETP Lett. {\bf 23}, 338(1976); M.Yu.Khlopov, Sov.J.Nucl.Phys. {\bf 28}, 583(1978).
    
\bibitem{fb0} M.A. Shifman and M.B. Voloshin, Yad. Fiz. {\bf 45}, 463 (1987).

\bibitem{fb1} M. Suzuki, Phys. Lett. B {\bf 162}, 392 (1985).

\bibitem{fb2} M. Neubert, Phys. Rept. {\bf 245}, 259 (1994).

\bibitem{fb3} M. A. Ivanov, Y. L. Kalinovsky, P. Maris and C. D. Roberts, Phys. Lett. B {\bf 416}, 29 (1998).

\bibitem{fb4} X. H. Guo and M. H. Weng, Eur. Phys. J. C {\bf 50}, 63 (2007).

\bibitem{fb5} S. Narison, Phys. Lett. B {\bf 198}, 104 (1987).

\bibitem{fb6} W. Lucha, D. Melikhov and Silvano Simula, Phys. Rev. D {\bf 88}, 056011 (2013).

\bibitem{DELPHI} J. Abdallah {\it et al.}, (DELPHI Collaboration), Eur. Phys. J. C {\bf 33}, 213 (2004).

\bibitem{HQET} C. Albertus, E. Hernandez, J. Nieves and J. M. Verde-Velasco, Phys. Rev. D {\bf 71}, 113006 (2005).

\end{thebibliography}
\end{document}